# Information-theoretic interpretation of tuning curves for multiple motion directions

Wentao Huang[1], Xin Huang[2] and Kechen Zhang[1]

[1]Department of Biomedical Engineering, Johns Hopkins University School of Medicine, Baltimore, Maryland 21205

[2]Department of Neuroscience, School of Medicine and Public Health, McPherson Eye Research Institute, University of Wisconsin-Madison, Madison, Wisconsin 53705

*Abstract*—We have developed an efficient information-maximization method for computing the optimal shapes of tuning curves of sensory neurons by optimizing the parameters of the underlying feedforward network model. When applied to the problem of population coding of visual motion with multiple directions, our method yields several types of tuning curves with both symmetric and asymmetric shapes that resemble what have been found in the visual cortex. Our result suggests that the diversity or heterogeneity of tuning curve shapes as observed in neurophysiological experiment might actually constitute an optimal population representation of visual motions with multiple components.

*Keywords—Shannon mutual information; Fisher information; unsupervised learning; area MT; bidirectional visual motion; side bias*

## I. Introduction

Sensory neurons respond selectively to specific features of the input stimuli and a basic description of this selectivity is a tuning curve, which quantifies how the average firing rate of a neuron depends on a given stimulus variable. In this paper we focus on the middle temporal (MT) area of the primates where neurons are highly selective to the direction of visual motion. The tuning curve of a typical MT neuron has a Gaussian shape with a single preferred direction [1]. When multiple motion directions are present simultaneously, the responses of MT neurons are more complex and more interesting, and the study of their response properties may help reveal how the biological visual system extracts useful motion information to segment moving objects.

The basic experimental fact relevant to our modeling project involves two sets of overlapping random dot stimuli moving in two separate directions [2]. Primate observers are able to segregate the two direction components and perceive motion transparency. As shown in Fig. 1 the responses of MT neurons have four general types. The first type is vector averaging (Fig. 1A), where the responses to the bidirectional stimuli are well approximated by averaging the responses to each of the two motion components. The next two types also have single-peaked tuning curves but the peak position is biased either to the right (i.e. the clockwise side) or to the left (counterclockwise side) from the midline (Fig. 1B and 1C).

The fourth type has double-peaked responses although the average of the responses to the two component directions contains only a single peak (Fig. 1D). The heights of the two peaks may be unequal. This type is qualitatively different from the first three types which all exhibit a single peak. Due to the symmetry of the stimulus setup, one would expect that the responses should also be symmetric, such as the tuning curve in Fig. 1A which is essentially symmetric. The side-bias types in Fig. 1B and 1C are unexpected because there is no obvious theoretical justification why directional symmetry should be broken here. A question naturally arises as to whether the asymmetric shape of these neurons may have some special functional meaning or computational advantages in the processing of visual motion with multiple components.

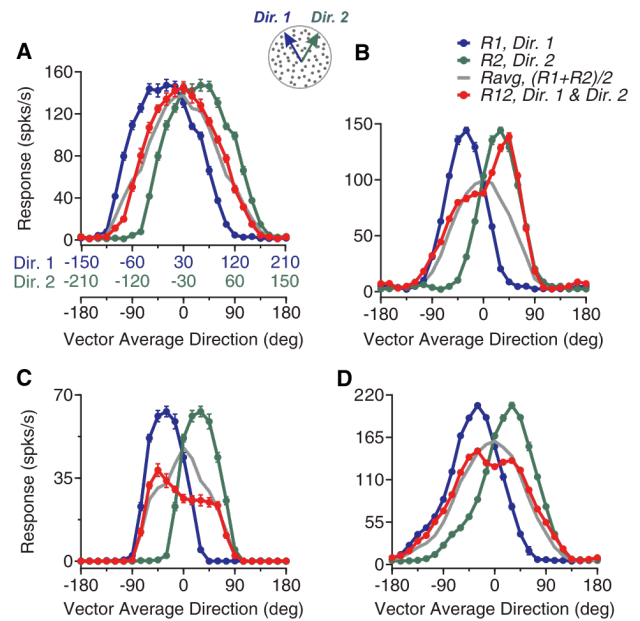

**Figure 1.** Four representing MT neurons showing the basic types of responses to random-dot stimuli with two motion directions. The angle between the two motion directions was always fixed at 60° while their average direction was varied systematically; the elicited neural responses are shown by the red curves. The responses to a single motion component without the presence of the other are also shown (blue or green), together with their averages (gray). Adapted from [2].

Since the visual system might be evolved to represent natural stimuli in an efficient manner [3]-[5], we have compared the empirical tuning curve shapes in Fig. 1 against the theoretically optimal shapes based on optimizing population coding for random motion stimuli. Our result shows an intriguing similarity between theory and experiment suggesting a novel information-theoretic interpretation of the experimental data.

## II. MODEL AND COMPUTATIONAL METHOD

### A. Maximizing Shannon Mutual Information

In this paper we optimize the population coding of MT neurons by maximizing Shannon mutual information between the stimuli and the neural responses. To evaluate mutual information efficiently for a population of neurons, we use an asymptotic formula that relates Shannon mutual information to the Fisher information matrix determinant, which tends to be much easier to optimize. This formula works extremely well for large populations of neurons, avoiding the problem of exponential explosion associated with evaluating mutual information by summing over all possible combinations of responses from all the cells in the population. Although Monte Carlo method could be used [6], its stochastic error is not suitable for the optimization problem in this paper. Another important benefit of the approach is that optimizing population distribution becomes a convex problem that guarantees a unique solution without local optima, allowing efficient numerical computation [7]. The Shannon mutual information $I$ between the stimuli and the neural responses has the following asymptotic formula in the limit of large neural population [7]:

$$I = \left\langle \ln \frac{p(\mathbf{x}|\mathbf{r})}{p(\mathbf{x})} \right\rangle_{\mathbf{r},\mathbf{x}} \simeq \frac{1}{2}\left\langle \ln\left(\det\left(\frac{\mathbf{G}(\mathbf{x})}{2\pi e}\right)\right)\right\rangle_{\mathbf{x}} + H \quad (1)$$

where each stimulus is specified as a vector $\mathbf{x} = (x_1,\cdots,x_K)^T$, the stimulus entropy is given by $H = -\langle \ln p(\mathbf{x})\rangle_{\mathbf{x}}$ the responses (spike counts) of a population of $N$ neurons are described by $\mathbf{r} = (r_1,\cdots,r_N)^T$, det denotes matrix determinant, and

$$\mathbf{G}(\mathbf{x}) = \mathbf{J}(\mathbf{x}) - \frac{\partial^2 \ln p(\mathbf{x})}{\partial \mathbf{x} \partial \mathbf{x}^T}, \quad (2)$$

with

$$\mathbf{J}(\mathbf{x}) = \left\langle \frac{\partial \ln p(\mathbf{r}|\mathbf{x})}{\partial \mathbf{x}} \frac{\partial \ln p(\mathbf{r}|\mathbf{x})}{\partial \mathbf{x}^T} \right\rangle_{\mathbf{r}|\mathbf{x}}, \quad (3)$$

being the Fisher information matrix. Here $\langle\cdot\rangle_{\mathbf{r},\mathbf{x}}$, $\langle\cdot\rangle_{\mathbf{r}|\mathbf{x}}$ and $\langle\cdot\rangle_{\mathbf{x}}$ denote the expectations with respect to the probability density functions $p(\mathbf{r},\mathbf{x})$, $p(\mathbf{r}|\mathbf{x})$ and $p(\mathbf{x})$, respectively. Assuming statistically independent Poisson noises of different neurons, we have $p(\mathbf{r}|\mathbf{x}) = \prod_{n=1}^{N} p(r_n|\mathbf{x},\theta_n)$ where vector $\theta_n$ represents the tuning parameters for the $n$-th neuron. To evaluate the Fisher information, we sum over discrete populations of neurons with distinct tuning parameters:

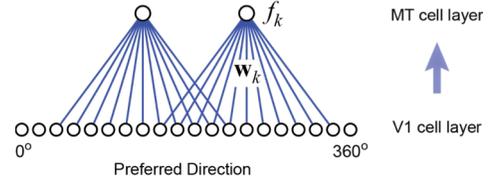

**Figure 2**. Feedforward network model for MT cell responses with inputs coming from the primary visual cortex V1. The cells are arranged according to their preferred directions of motion. The connection weights between the two layers along with other parameters are optimized by information maximization.

$$\mathbf{J}(\mathbf{x}) = N \sum_{k=1}^{K_1} \rho_k \left\langle \frac{\partial \ln p(r|\mathbf{x},\theta_k)}{\partial \mathbf{x}} \frac{\partial \ln p(r|\mathbf{x},\theta_k)}{\partial \mathbf{x}^T} \right\rangle_{r|\mathbf{x}}, \quad (4)$$

where we assume that there are $K_1$ subpopulations of identical neurons, and $\rho_k \geq 0$ is the population density for the $k$-th subpopulation, with $\sum_{k=1}^{K_1} \rho_k = 1$. We can prove that finding the population density $\rho_k$ that maximizes the asymptotic form of Shannon mutual information is a convex optimization problem, which ensures that efficient numerical optimization is tractable for large neural populations [7], [8].

### B. Network Architecture and Optimization Method

We use a feedforward network to model the information processing from the primary visual cortex (V1) to area MT (Fig. 2). We denote each stimulus as vector $\mathbf{s} = (s_1, s_2, \cdots, s_{i_{max}})^T$ whose $i$-th element indicate the intensity at direction $\theta_i$ ($i = 1, 2, ..., i_{max}$) and $i_{max} = 12$. For instance, when the stimulus has a single movement direction, all elements of vector $\mathbf{s}$ are 0 except the one corresponding to the movement direction. When the stimulus has two movement components, all elements of vector $\mathbf{s}$ are 0 except the two elements that correspond to the two directions. These stimuli were first processed with a population of $M = 24$ model V1 cells, each with a preferred direction at one of the $K$ directions around the circle. The directional tuning curve for the $m$-th model V1 neuron is described by a von Mises function:

$$v_m(\theta_i) = \frac{1}{1-b}\left(\exp\left(\frac{\cos(\theta_i - c_m)-1}{\sigma^2}\right) - b\right) \quad (5)$$

where parameter $b = \exp(-2/\sigma^2)$ is chosen so that the response range is normalized between 0 and 1, parameter $c_m = 15° \times m$ ($m = 1, 2, ..., M$) is the directional preference of this V1 neuron, and parameter $\sigma = \pi/2$ is the directional tuning width. The trial-by-trial linear response of each model V1 neuron is computed by

$$x_m = \sum_i v_m(\theta_i)s_i. \quad (6)$$

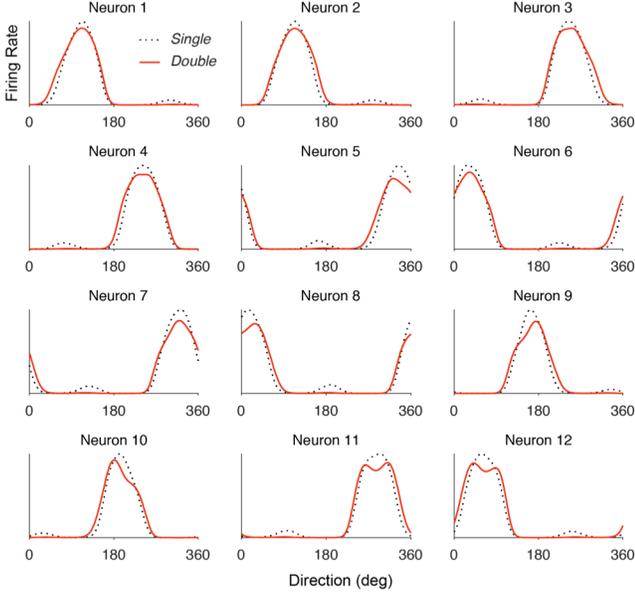

**Figure 3.** The optimal response tuning curves of 12 model MT neurons elicited by bidirectional stimuli separated by 60° (red curves). The horizontal axis refers to the direction of the midline of the two directions. For comparison, the responses to single movement direction are shown as black dotted curves.

Now the input of all MT neurons is defined by the vector $\mathbf{x} = (x_1, x_2, ..., x_M)^T$. The raw response of the $k$-th MT neuron before divisive normaliztion is given by

$$f_k(\mathbf{x}) = \frac{A}{1 + \exp(-\mathbf{w}_k^T \mathbf{x} + b_k)} \quad (7)$$

where $A$ is a constant, $\mathbf{w}_k = (w_{1,k}, w_{2,k}, ..., w_{M,k})^T$ is the weight vector, and $b_k$ is a threshold parameter ($k = 1, 2, ..., K$). To obtain the final tuning curve of the $k$-th MT neuron, the raw responses are squared and normalized as follows

$$\hat{f}_k(\mathbf{x}) = \frac{f_k^2(\mathbf{x})}{\sum_{k'} f_{k'}^2(\mathbf{x}) + \varepsilon}, \quad (8)$$

where $\varepsilon = 0.1$ is a constant.

To maximize mutual information $I$ while constraining the total energy, we minimize the following objective function [7], [8]:

$$Q\left[\{\rho_k, \mathbf{w}_k, b_k\}\right] = \left\langle -\frac{1}{2} \ln(\det(\mathbf{G}(\mathbf{x}))) + \lambda \sum_{k=1}^{K_1} \rho_k \hat{f}_k(\mathbf{x}) \right\rangle_\mathbf{x}, \quad (9)$$

where constant $\lambda = 10$, and the parameters to be optimized include the population density $\rho_k$, the connection weights $\mathbf{w}_k$ and the threshold $b_k$ ($k = 1, 2, ..., K$). In our simulations described below, $K = 12$ types of MT cells were used. The objective function in Eq. (9) was optimized using the method detailed in [8].

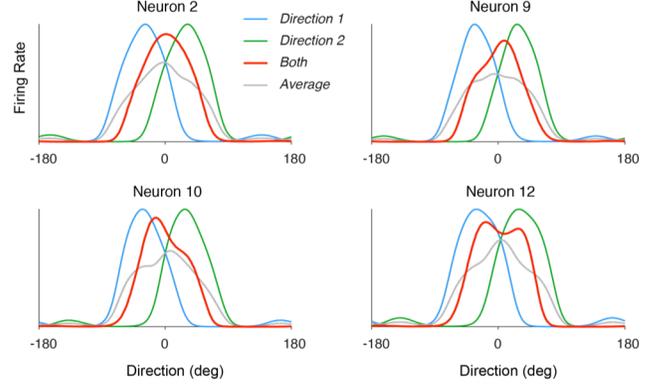

**Figure 4.** Four examples of cell types taken from Fig. 3 are plotted in a format similar to Fig. 1. The responses to bidirectional stimuli are shown (red), with the horizontal axis indicating the direction of the midline of the two components relative to the preferred direction of the neuron. The responses to each of the two components presented alone (blue or green) are also plotted together with their averages (gray).

The stimuli were generated randomly with either one motion direction or two motion directions. The training samples include 180,000 single direction stimuli, and 36,000 bidirectional stimuli with random angular separation given by $15^\circ \times n$ ($n = 1, 2, ..., 12$).

### III. SIMULATION RESULTS

We have found several different types of tuning curves in the optimized MT model neurons (Fig. 3). In this example a total of $K = 12$ MT cells were used in the optimization procedure. More rigorously speaking, each cell is actually a representative of a subpopulation of identical cells or cells with identical tuning curves. There is no other requirement besides that the subpopulation should be large enough. To compare the theoretical results with the experimental data in Fig. 1, we picked four cells from Fig. 3 and plotted the theoretical curves in the same format (Fig. 4). Comparing Fig. 1 and Fig. 4, we see a general resemblance with different types of tuning curves. In Fig. 4, Neuron 2 shows symmetric tuning curve for bidirecdtional stimuli although the responses are not as close to the average as its experimental counterpart. The asymmetric tuning curves of Neuron 9 and Neuron 10 show unmistakable side biases. Neuron 12 clearly has a tuning curve with two peaks. Compatible with the experimental data, the responses to bidirectional stimuli tend to have lower peak rate and wider tuning curve width than the stimuli with a single movement direction, although the widths are not as wide as the experimental data. We have also varied the angle of separation between the two motion directions and found that the asymmetric shapes had a tendency to be biased toward the same side across different separation angles, which is consistent with the experimental finding. So the theoretical optimal tuning curves are qualitatively similar to what have been observed in experiment.

In the experimental data, the most abundant type is the single-peaked average type (Fig. 1A), which constitutes ~40% of all cells. The clockwise side-bias type and the counterclockwise side-bias type each constitutes ~20% of cells while the two-peaked type also constitutes ~20% of cells [2]. In the theoretical simulation (Fig. 3), the symmetric single-peaked type constitutes 50% of cells, whereas all other three types each has about 17% of cells. So the theoretical optimal population distribution is roughly compatible with the population distribution found in neurophysiological experiment.

## IV. Discussion

We have developed an efficient computational method for optimizing the Shannon mutual information between random visual motion stimuli and the elicited responses of a population of MT neurons with independent Poisson spikes. The parameters optimized here include the connection weights between a layer of MT units and a lower layer of units with V1-like directional responses. The model MT neurons obtained after the training optimization process show various tuning curves, including both symmetric and asymmetric shapes. This result suggests that a subpopulation of asymmetric tuning curves, in combination with another subpopulation of symmetric tuning curves, may actually form an optimal solution for encoding motions with multiple components.

It is worth mentioning that in the optimization procedure, no information about the experimental tuning curves was ever used. In other words, our method is not equivalent to curve fitting, and the theoretical MT tuning curves may potentially take any form. We only optimize the mutual information between the stimuli and the neural responses by an unsupervised learning process. After learning, different types of tuning curves that resemble the experimental data emerge spontaneously. This result may suggest that the real population coding in area MT might be optimized for dealing with visual motions with multiple components.

This work could be readily generalized in several directions. One could explore the statistics of natural motions in order to better constrain the stimuli used during the unsupervised learning process. More realistic statistics of motion stimuli should be incorporated into the set of our training stimuli [9], [10]. One could also incorporate additional parameters into the model, such as illuminance, movement speed, and binocular disparity. Such investigation is expected to shed light on the relationship between the response properties of MT neurons and the statistical regularities of movements in the natural world while producing explicit MT models with realistic tuning curve shapes. Since the information-theoretic method described here is quite general, it might be applicable to related models of sensory neurons in other visual areas or other sensory modalities.

ACKNOWLEDGMENT

Supported by grants NIH R01 EY022443 and NIH R01 DC013698.